\documentclass[useAMS,usenatbib,usegraphicx]{mn2e}

\title[Gas and stellar metallicities in HII galaxies]{Gas and stellar metallicities in HII galaxies}

\author[P. Westera et al.]{P.~Westera$^{1,2}$\thanks{E-mail: pieter.westera@ufabc.edu.br, pieter@on.br},
F.~Cuisinier$^{3}$\thanks{In memoriam (1969-2011)},
D.~Curty$^{2}$\thanks{E-mail: curty@on.br}
and R.~Buser$^{4}$\thanks{E-mail: Roland.Buser@unibas.ch}\\
$^{1}$Universidade Federal do ABC, Rua Santa Ad\'{e}lia, 166, 09.210-170, Santo Andr\'{e} - SP, Brazil\\
$^{2}$Observat\'{o}rio Nacional, Rua Jos\'{e} Cristino, 77, 20.921-400, Rio de Janeiro - RJ, Brazil\\
$^{3}$Observat\'{o}rio do Valongo/UFRJ, Ladeira do Pedro Ant\^{o}nio, 43, 20.080-090 Rio de Janeiro - RJ, Brazil\\
$^{4}$Universit\"{a}t Basel, Departement Physik, Klingelbergstrasse 82, 4056 Basel, Switzerland}
\begin{document}

\date{Accepted 1988 December 15. Received 1988 December 14; in original form 1988 October 11}

\pagerange{\pageref{firstpage}--\pageref{lastpage}} \pubyear{2002}

\maketitle

\label{firstpage}

\begin{abstract}
   We examine the gas and stellar metallicities in a sample of
   HII galaxies from the Sloan Digital Sky Survey,
   which possibly contains the largest homogeneous sample of HII
   galaxy spectra to date.

   We eliminated all spectra with an insufficient
   signal-to-noise ratio, without strong emission lines,
   and without the [OII]~$\lambda$3727~\AA\ line, which is
   necessary for the determination of the gas metallicity.
   This excludes galaxies with redshift $\stackrel{_<}{_\sim} 0.033$.
   Our final sample contains $\sim$700 spectra of HII galaxies.

   Through emission line strength calibrations and a detailed
   stellar population analysis employing evolutionary stellar
   synthesis methods, which we already used in previous works,
   we determined the metallicities of both the gas and the
   stellar content of these galaxies.

   We find that in HII galaxies up to stellar masses of
   $5\cdot10^9~M_{\odot}$, enrichment mechanisms do not vary
   with galactic mass, being the same for low- and high-mass
   galaxies on average.
   They do seem to present a greater variety at the high-mass end,
   though, indicating a more complex assembly history for high-mass
   galaxies.
   In around 23 per cent of our HII galaxies we find a metallicity
   decrease over the last few Gyr.
   Our results favour galaxy evolution models featuring constantly
   infalling low-metallicity clouds that retain part of the galactic
   winds.
   Above $5\cdot10^9~M_{\odot}$ stellar mass, the retention of high
   metallicity gas by the galaxies' gravitational potential dominates.
\end{abstract}


\section{Introduction}

   HII galaxies are characterised by prominent emission lines.
   In fact, they are defined as having strong H$\beta$ and
   OIII lines, but other ionised hydrogen (H$\alpha$, H$\gamma$,
   etc.) and high excitation metallicity lines (OII, NII, and others)
   are strong too.
   These emission lines are driven by photoionisation of interstellar
   gas by hot massive stars from young stellar populations
   \citep{sargent,french}.
   The star forming activity in HII galaxies is so strong that
   it can certainly not have been maintained at its present level
   during a Hubble time \citep*[see e.g.][]{searle2}.
   In fact, they are the most extreme case of star forming galaxies,
   showing the highest excitation emission lines, and thus the highest
   (relative) star formation rates (SFRs). \\
   They are also the lowest metallicity galaxies of the interstellar
   medium, with metallicities around one tenth solar
   \citep*[][and others]{vilchez,perezmonteroa,kniazev,kehrigvilchez,izotov,haegele,perezmonteroc}. \\
   It is now clear that the hypothesis of HII galaxies as galaxies
   experiencing their first star formation burst can be ruled out, and
   that their stellar content is predominantly old, older than one Gyr
    \citep*[][and others]{raimann,cidfernandes,kong,westera_04,cidasari,asari,hoyos},
   Nevertheless, both the high relative SFRs and the low metallicities
   indicate that HII galaxies are among the least evolved galaxies
   in existence, which assigns to them a special role as fossil record
   of galaxy evolution \citep{lequeux}.
   They are, therefore, ideal objects for chemical evolution studies.

   Several authors have developed theoretical evolutionary models of
   dwarf irregular galaxies, of which HII galaxies are a sub-category.
   Some of these models predict the galaxies' present-day stellar - and gas
   metallicities, among other properties.
   In these models, several chemical enrichment processes are
   identified. \\
   \citet{maclow} developed single-phased hydrodynamical models for
   dwarf galaxies.
   They model the effects of repeated supernova type II (SNII) explosions
   from starbursts on the interstellar medium, that is,
   the enrichment by the ejected metals on the one hand, and
   the gas loss by supernova winds on the other hand, taking into
   account the gravitational potential of their dark matter haloes.
   They find that, in galaxies with gas masses below $10^6~M_{\odot}$,
   most of the gas is blown away, and in galaxies with masses between
   $10^7~M_{\odot}$ and $10^9~M_{\odot}$, mainly the newly-formed
   metal-rich gas is blown away, whereas the already present, metal-poor,
   gas is retained.
   They do not calculate the final (present-day) metallicities of their
   model galaxies, but it is obvious that they remain low-metallicity,
   at least the ones with gas masses below $10^9~M_{\odot}$. \\
   \citet{recchimatta,recchimattb} simulate IZw18, one of the
   lowest-metallicity galaxies known, using a single star formation
   burst model and a model with a doubly peaked star formation
   history.
   They use a lower SNII heating efficiency than MacLow et al.
   in their models, but include type Ia supernovae (SNeIa),
   for which they use a higher heating efficiency.
   As a consequence, alpha elements are ejected less efficiently
   than in MacLow et al.'s model, resulting in higher metallicities.
   The iron-peak elements, which are produced in SNeIa,
   on the other hand, are still ejected, causing high [$\alpha$/Fe]
   ratios.
   The (gas) metallicity in their single burst model remains very low,
   [O/H$]\stackrel{_<}{_\sim}-1.3$, depending on the galaxy mass.
   In their (more likely) double burst model, [O/H] can reach values up
   to -1, depending on the details of the models, that is, the stellar
   yields used, the stellar initial mass function (IMF), the time between
   the bursts, the duration of the second burst, the gas density and -
   metallicity in the star forming region, and the total mass of the
   stars formed in the second burst.
   The final metallicities of the double burst model galaxies are
   between 0.6 and 1 dex higher than after the first burst. \\
   \citet{tenoriotagle} include photoionisation and cluster wind in
   their two-dimensional hydrodynamic calculations.
   They identify two mechanisms:
   The storage of clouds into a long-lasting ragged shell inhibiting
   the expansion of the thermalised wind, and the steady filtering of
   the shocked wind gas through channels carved within the cloud layer.
   They conclude that both processes must be at work in HII galaxies.
   Unfortunately, they make no prediction about the present-day metallicity
   of their model galaxies. \\
   In a series of papers, Recchi and collaborators use a new generation of
   chemo-dynamical models, some of which include infalling clouds.
   The first two, \citet{recchimattc} and \citet{recchihenslera}, are
   aimed at reproducing the properties of two specific objects,
   IZw18 and NGC1569, respectively.
   The third one, \citet{recchihenslerb} investigates the influence of
   several factors, such as the cloudiness of the gas distribution,
   the IMF slope, and stellar yields.
   They find that models with continuous low level star formation periods
   in the past followed by a quiescent phase and a recent, stronger burst
   best reproduce the chemical properties of the studied galaxies.
   They also find that, for a homogeneous gas distribution, metals get
   blown away by galactic winds, but cool infalling intergalactic clouds
   can hamper these galactic winds.
   The final metallicities in the their various models vary from
   -2.5 to -0.8.

   Although metallicity determinations of the various components of
   HII galaxies cannot discriminate, how much each of these mechanisms
   contribute to the chemical evolution, they can quantify the
   metallicity at various stages of a galaxy's evolution, i.e.
   at the time of the formation of the old ($>1$ Gyr) stellar populations
   and at present, and thereby help to rule out some of the evolutionary
   scenarios, and support others.
   HII galaxies have the advantage that it is possible to determine
   independently the metallicities of the young and the old populations
   from their spectra.
   As the young populations were formed recently out of the same
   gas that is responsible for the emission lines, their metallicity
   can be assumed to be more or less the same as the one of the
   gas, which can be determined from the emission line strengths.
   The metallicity of the old populations, on the other hand, can be
   determined from the continuum and absorption features through
   population synthesis. \\
   The main goal of the present work is to measure the chemical enrichment
   in a homogeneous sample of HII galaxies, and examine possible
   trends with galactic properties such as total mass.
   This way, we can evaluate the different models and make statements
   on the importance of the various enrichment mechanisms.

   Until recently, the discovery of HII galaxies was limited to the visual
   inspection of objective-prism surveys, introducing ill-controlled
   biases and selection effects
   \citep*[for compilations, see][and references therein]{terlevich,kehrig}. \\
   The Sloan Digital Sky Survey \citep*[SDSS;][]{york} presents for the
   first time a comprehensive database of galactic spectra.
   Data release 7 \citep{abazajian} contains over 900'000 galaxy spectra
   that have been selected on clear magnitude limiting criteria.
   In an unprecedented manner, the SDSS allows to study star forming emission
   line galaxies based on clear and quantitative criteria,
   and not mere visual inspection.

   In this work, we define clear emission line strength criteria to
   distinguish galaxies containing strongly excited gas from other
   emission line galaxies.
   We only include spectra that contain the
   [OII]~$\lambda$3727~\AA\ emission line, as this line is indispensable
   for an adequate determination of the gas metallicity.
   This criterion excludes galaxies with redshift
   $\stackrel{_<}{_\sim} 0.033$.
   This way, we selected a homogeneous sample of $\sim 700$ HII galaxy
   spectra of high quality from the SDSS. \\
   In this sample, we determine the metallicities of both the gas and
   the old stellar content using emission line calibration
   and stellar population synthesis methods, respectively.
   It is the first time an independent metallicity determination
   of the gas and the stars is performed on a clearly defined
   HII galaxy sample.

   The layout of this article is the following:
   In Sect.~\ref{spectra}, the catalogue of spectra analysed in this work
   is presented.
   Sect.~\ref{method} gives a detailed description of the method we used
   to analyse the spectra.
   Subsection \ref{OHgas} describes, how we determined the gas
   metallicity, and Subsection \ref{FeHstars} is dedicated to the
   determination of the stellar metallicities.
   The main results are given in Sect.~\ref{results} and discussed
   in Sect.~\ref{discussion}.
   A summary and the main conclusions can be found in Sect.~\ref{conclusions}.

\section{The data base}
\label{spectra}

   \begin{figure}
    \includegraphics[width=\columnwidth]{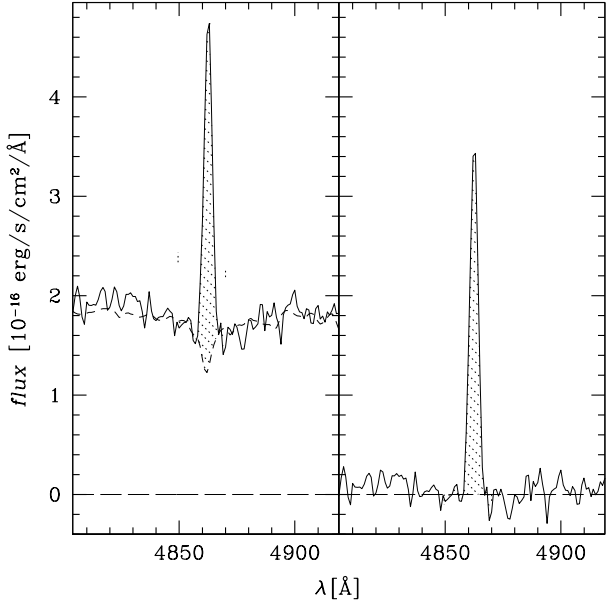}
    \caption{Illustration of the emission line strength
       measurement (here H$\beta$ in 51957-0273-418).
       In the left panel, the solid line represents the observed
       spectrum (rebinned to the BC03 wavelength grid), whereas
       the short-dashed line shows the best fit using the BC03
       high resolution library.
       The shaded region between these two lines shows the area
       used to calculate the emission-line strength.
       The right panel shows the emission line after subtracting
       the best fit.
       The shaded area corresponds to the shaded area between
       spectrum and best fit from the left panel.}
    \label{subtraction}
   \end{figure}
   \begin{figure}
    \includegraphics[width=\columnwidth]{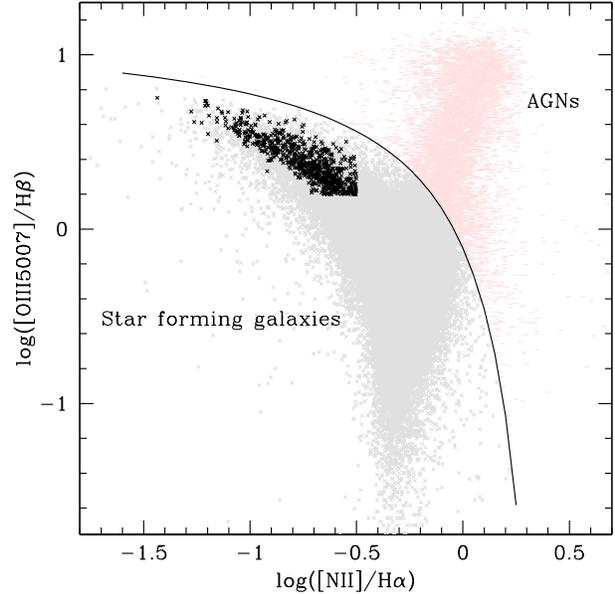}
    \caption{Excitation diagram for all galaxies of our sample
       (black crosses).
       The grey crosses represent the test sample, whereas
       galaxies identified as AGNs are shown as
       dashes.
       The solid line shows models from \citet{kewley}
       used to separate the AGNs from star forming galaxies.}
    \label{seagull}
   \end{figure}

   For this work, we used spectra from the SDSS data release 7,
   which have a signal-to-noise of $S/N>4$ per pixel at $m_g=20.2$, but
   usually much higher, and cover a wavelength range of 3800 to 9200~\AA.
   Most important of all, it contains a large number of galaxy spectra,
   that is around 900'000.

   Any analysis however relies on previous clear criteria to define what
   HII galaxies are.
   First, we rejected all spectra whose wavelength ranges do not include
   the region from 3677~\AA\ to 3775~\AA, which is needed to measure the
   [OII]~$\lambda$3727~\AA\ emission line strength,
   used in many strong line methods to determine the gas metallicity.
   This excludes galaxies with redshifts below 0.033.
   Another prerequisite for measuring this line strength is a sufficient
   signal-to-noise ratio in this wavelength region.
   We only kept spectra with $S/N_{{\rm [OII]3727}}\geq 2.8$.
   This criterion guarantees sufficiently low noise for the entire
   spectral range used in this work, since at longer wavelengths, the
   signal-to-noise ratio of the spectra is higher than around the
   [OII]~$\lambda$3727~\AA\ line, i.e. $\stackrel{_>}{_\sim}8$ in $g$. \\
   The spectra were then corrected for Galactic (foreground) gas extinction
   using the values given in the SDSS database, which were derived from
   the \citet*{schlegel} reddening maps.
   Using these extinction values
   and the Galactic extinction law of \citet{fitzpatrick},
   we dereddened the spectra.
   After that, we deredshifted them, and measured the emission line
   strengths. \\
   In order to properly determine the emission line strengths, we first
   had to remove the contribution from the absorption lines of the
   underlying stellar populations.
   This was done by subtracting high resolution spectra representing
   the stellar continuum from the empirical spectra as illustrated in
   Fig~\ref{subtraction}.
   These high resolution spectra were compiled by fitting the spectra of
   composite stellar populations, made up of three single stellar
   populations (SSPs), a young ($\leq 10$~Myr), an intermediate-age
   (20~--~500~Myr), and an old one (5~Gyr), from the ``BC03''
   integrated spectral energy distribution (ISED) library
   to the continua of the galaxy spectra.
   The ``BC03'' library was produced using the Bruzual and Charlot (2003)
   Galaxy Isochrone Spectral Synthesis Evolution Library (GISSEL) code
   \citep{charlot_91,bruzual_93,bruzual_03} implementing the Padova 1995
   isochrones \citep{fagotto,girardi} combined with the STELIB
   \citep{leborgne} stellar library.
   Since it is made up of empirical high resolution spectra, the ``BC03''
   library reproduces absorption line shapes well, and is ideal for
   this purpose.
   Table~\ref{lines} lists all the lines we measured. \\
   However, incomplete coverage of the stellar parameter space and
   calibration uncertainties of the STELIB library result in systematic
   errors in the overall spectral shapes.
   Therefore the ``BC03'' library is not a good choice for determining
   the properties of the stellar content from a full spectral fit.
   For the stellar population synthesis, we perform another fit using
   a different library, described in the next section. \\
   \begin{table}
   \begin{center}
      \caption{Emission lines measured in this work.}
      \label{lines}
         \begin{tabular}{ll}
            \hline
   line & central wavelength \\
            \hline
   {}[OII]~$\lambda$3727~\AA & 3727 \AA \\
   H$\epsilon$ & 3970 \AA \\
   H$\delta$ & 4102 \AA \\
   H$\gamma$ & 4340 \AA \\
   {}[OIII]~$\lambda$4363~\AA & 4363 \AA \\
   {}[HeII]~$\lambda$4686~\AA & 4686 \AA \\
   H$\beta$ & 4861 \AA \\
   {}[OIII]~$\lambda$4959~\AA & 4959 \AA \\
   {}[OIII]~$\lambda$5007~\AA & 5007 \AA \\
   {}[HeI]~$\lambda$5876~\AA & 5876 \AA \\
   {}[OI]~$\lambda$6300~\AA & 6300 \AA \\
   {}[SIII]~$\lambda$6312~\AA & 6312 \AA \\
   {}[NII]~$\lambda$6548~\AA & 6548 \AA \\
   H$\alpha$ & 6563 \AA \\
   {}[NII]~$\lambda$6584~\AA & 6584 \AA \\
   {}[SII]~$\lambda$6717~\AA & 6717 \AA \\
   {}[SII]~$\lambda$6731~\AA & 6731 \AA \\
   {}[OII]~$\lambda$7319~\AA & 7319 \AA \\
   {}[OII]~$\lambda$7330~\AA & 7330 \AA \\
            \hline
         \end{tabular}
   \end{center}
   \end{table}
   Finally, we corrected the spectra for internal gas extinction
   using the \citet{calzetti} attenuation law.
   The extinction constants
   $E(B-V)=0.44\times6.60\log(I_{{\rm H}\alpha}/I_{{\rm H}\beta}/2.87)/4.04$
   were estimated from the
   ${\rm H}\alpha /{\rm H}\beta$ Balmer decrements following
   \citet*{calzetti}, adopting intrinsic ratios
   $I_{{\rm H}\alpha}/I_{{\rm H}\beta}=2.87$ \citep{osterbrock}. \\
   The factor of 0.44 stems from the correction
   for differential extinction between the stellar
   populations and the gas following \citet*{calzettikin}.
   As our sample consists of HII galaxies only, which is a subclass of
   starburst galaxies, using this factor is justified.
   The emission line strengths were dereddened as well, using the same
   $E(B-V)$ values but without multiplying them by 0.44.

   To make sure our sample contains only HII galaxies, we limited it
   to galaxies with strong emission lines at a high excitation level,
   by applying the two criteria:
   \begin{equation}
    {\rm log}\left(\frac{{\rm[OIII]}~\lambda5007~\rm{\AA}}{{\rm H}\beta}\right)\geq0.2\parallel{\rm log}\left(\frac{{\rm[NII]}}{{\rm H}\alpha}\right)\leq-0.5,
   \end{equation}
   whereas [NII] = [NII]~$\lambda$6548~\AA\ + [NII]~$\lambda$6584~\AA.
   Of course, these criteria are more restrictive than the traditional
   definition of HII galaxies as galaxies showing strong Balmer lines.
   In fact, as our criteria are designed to identify galaxies with gas at a
   high excitation level, our sample is biaised towards extreme case
   HII galaxies.
   As a consequence, we expect the galaxies of our sample to show typical
   properties of HII galaxies, but at a more pronounced level.
   For example, it consists of HII galaxies with low mass-to-light ratios
   and thus contains mainly dwarf galaxies.
   The reason for this approach, apart from emphasising on typical
   HII galaxy properties, is, to make sure that no ``contaminating''
   objects, such as LINERs, enter the sample.
   Then, we separated the star forming galaxies from AGNs, using the
   \citet{kewley} excitation line criterion,
   \begin{equation}
    {\rm log}\left(\frac{{\rm[OIII]}~\lambda5007~\rm{\AA}}{{\rm H}\beta}\right)<
    \frac{0.61}{{\rm log}({\rm[NII]}/{\rm H}\alpha)-0.47}+1.19,
   \end{equation}
   as shown in Fig.~\ref{seagull}. \\
   Our final sample contains 712 HII galaxy spectra.
   The relatively low number of galaxies compared to the total number
   of over 900'000 galaxies in the SDSS data release 7 is due to our high
   requirements towards the quality of the spectra, and to the criterion
   that the spectra must contain all the lines necessary for our study.
   This way, we guarantee that the sample is not contaminated, and that
   the obtained results are trustworthy.
   However, the sample is large enough to be statistically significant.

   We also defined a test sample, containing galaxies with some line
   emission, but not necessarily at a high level, that is H$\alpha$,
   H$\beta$, [OII]~$\lambda$3727~\AA, [OIII]~$\lambda$4959~\AA,
   [OIII]~$\lambda$5007~\AA, ${\rm[NII]}>0$.
   The reason for defining a test sample is, on the one hand,
   to identify properties typical of HII galaxies through comparison
   with the test sample, and, on the other hand, to check our method,
   by comparing the results for our test sample with the ones
   obtained in other studies using similar galaxy samples.
   As we needed to measure the [OII]~$\lambda$3727~\AA\ line strength
   for the test sample as well, it is also limited to galaxies with
   redshift $\stackrel{_>}{_\sim} 0.033$.
   To guarantee sufficient quality,
   we applied the signal-to-noise criterion $S/N_g\geq5$,
   where the $S/N_g$ values were taken from the SDSS database.
   Here too, we removed the AGNs using the \citet{kewley} criterion.
   This way, our test sample contains 74989 spectra.
   Given the criterion that the test sample galaxies show some line
   emission, it consists of star forming galaxies, probably mostly spirals.

   Fig.~\ref{seagull} shows both our sample of HII galaxies
   (in black), and the test sample (in grey) in a
   Baldwin--Phillips--Terlevich \citep*[BPT, ][]{BPT}
   classification diagramme.

\section{Method}
\label{method}

\subsection{Determination of the gas metallicities}
\label{OHgas}

   \begin{figure*}
    \includegraphics[width=\textwidth]{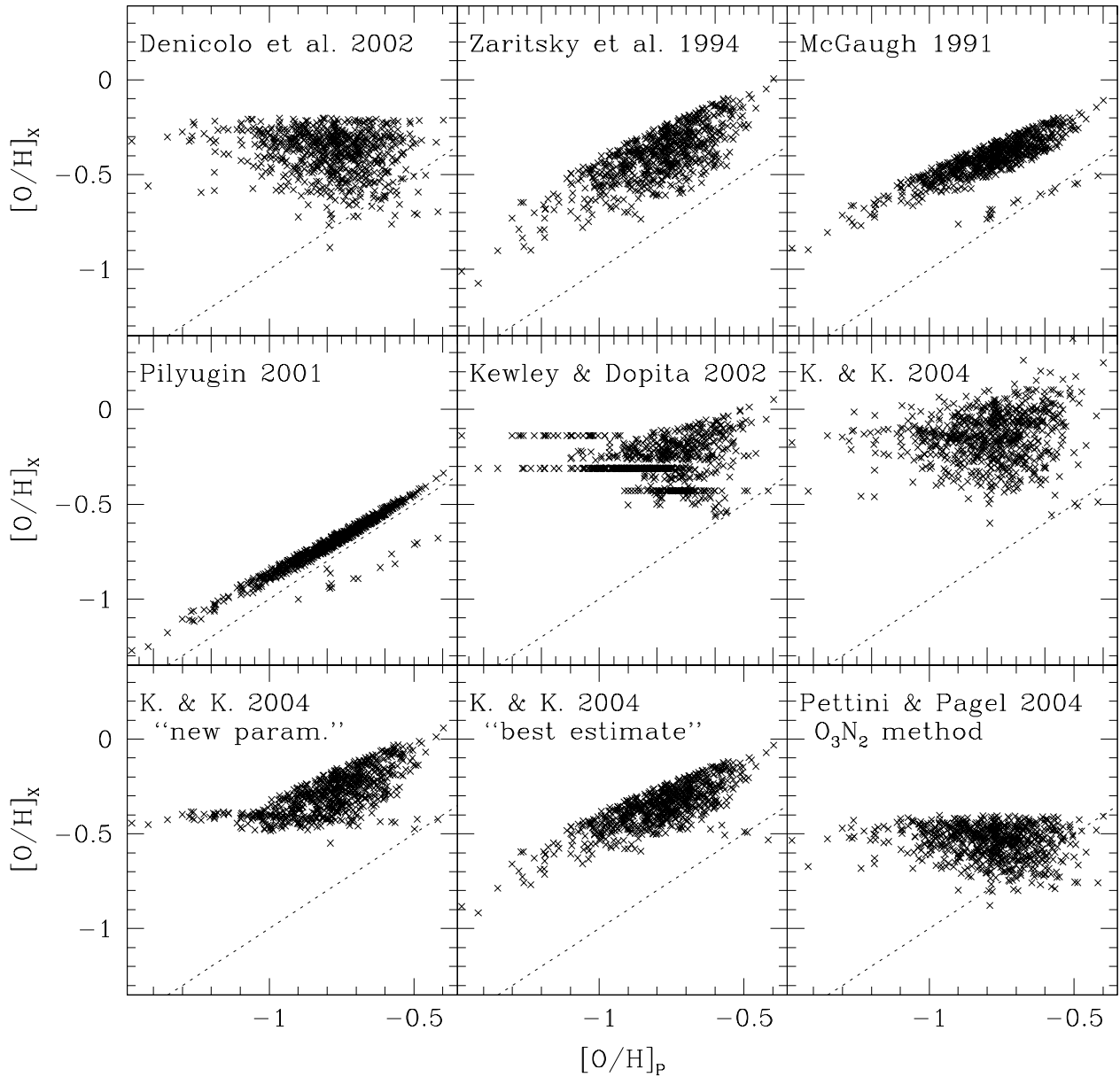}
    \caption{Comparison between the metallicities determined using
       the \citet{pilyuginthuan} method ($x$-axes) and the ones
       determined by nine other strong-line methods ($y$-axes).
       K. \& K. 2004 stands for \citet{kubilnickykewley}.
       The dashed lines show unity, [O/H]$_{\rm X}=$~[O/H]$_{\rm P}$
       For further details, see to the text.}
    \label{OHpvsotherOH}
   \end{figure*}

   We first measured the gas metallicity, which corresponds to the
   metallicity of the young stellar populations.
   In order to derive gas abundances from its line emission, hydrogen
   lines are needed, and the lines of at least one ion, generally oxygen,
   in its two dominant ionisation stages.
   Unfortunately, the electron temperature method using the
   [OIII]~$\lambda$4363~\AA\ emission line could not be used as this line
   and other auroral lines are much too weak to be measured with
   the necessary precision at the signal-to-noise ratios of the SDSS spectra.
   In most spectra, we do not even detect the
   [OIII]~$\lambda$4363~\AA\ line at all.
   Hence, we are restricted to strong line methods.
   The [OII]~$\lambda$7330~\AA\ line could be used, but it is too sensitive
   to the electron temperature and is possibly ``contaminated'' by
   recombination contributions. \\
   We compare the gas metallicities derived from 11 different
   strong line methods. \\
   A frequently used indicator is the [NII]~$\lambda$6548~\AA\ or the
   [NII]~$\lambda$6584~\AA\ line \citep{denicolo,pettinipagel}.
   However, since nitrogen is produced and destroyed during both
   primary and secondary nucleosynthesis, its abundance correlates
   in a non-evident way on (oxygen) metallicity.
   Therefore, it should be used only for rough estimates, to discriminate
   between the different branches of a multiple-valued method,
   for instance. \\
   One such multiple-valued indicator is the $R_{23}$ parameter \citep{pagel},
   which sums up the fluxes of the [OII]~$\lambda$3727~\AA\ and two strong
   [OIII] lines, at 4959~\AA\ and 5007~\AA.
   As $R_{23}$ is aproximately proportional to the oxygen abundance at 
   low metallicities, but decreases at high metallicities due to
   cooling, this indicator has a low - and a high metallicity branch.
   In spite of this double-valuedness, the $R_{23}$ parameter is used in many
   metallicity calibrations
   \citep{mcgaugh,zaritsky,pilyugina,pilyuginb,kewleydopita,kubilnickykewley,pilyuginthuan},
   some of which include a recipe, how to determine, on which branch
   a given galaxy lies, whereas others leave it up to the user to
   decide this.
   In one case, \citet{zaritsky}, a calibration is determined only for
   the high metallicity branch.
   Another point that has to be taken into account \citep[although in some methods
   this is ignored, i.e.][]{zaritsky} is, that $R_{23}$ depends not only
   on the metallicity of the ionised gas, but also on the hardness of the
   ionising radiation, which can be
   quantified by the ionisation parameter $q$, representing the
   ionising photon flux per unit area divided by the number density of
   hydrogen atoms.
   \citet[``new paramatrisation'' and ``best estimate'' methos]{kewleydopita,kubilnickykewley} use the [OIII]/[OII] ratio to
   calculate $q$ in an iterative way from theory, and then determine
   [O/H] from $R_{23}$ and $q$,
   whereas \citet{mcgaugh} uses the [OIII]/[OII] ratio directly in an
   empirical calibration without determining $q$.
   In \citet{pilyugina,pilyuginb,pilyuginthuan}, the so-called
   excitation or ionisation or, simply, P parameter is introduced
   for their empirical calibrations.
   It is defined as the [OIII]/([OII]+[OIII]) ratio, so the
   [OIII]/[OII] ratio is related to it, ${\rm [OIII]/[OII]}={\rm P}/(1-{\rm P})$.
   Recapitulating, in all of these methods the gas metallicity is
   calculated from the $R_{23}$ parameter and the [OIII]/[OII] ratio. \\
   Other strong line methods include \citet[{[NII]} method]{kubilnickykewley},
   combining the [NII]~$\lambda$6584~\AA\ line and the [OIII]/[OII] ratio,
   and \citet{pettinipagel}, employing the ratio between the
   [OIII]~$\lambda$5007~\AA\ and [NII]~$\lambda$6584~\AA\ lines.

   Fig.~\ref{OHpvsotherOH} shows the metallicities of our HII galaxy
   sample as determined by nine of the methods we studied, plotted
   against the ones using by the tenth method, \citet{pilyuginthuan},
   which is the one we ended up adopting.
   The only calibration not shown here due to space limitations is
   the \citet{pettinipagel} [NII] one, which is a vertically squeezed
   and shifted downward version of the \citet{denicolo} calibration
   (first panel). \\
   Most methods yield similar relative metallicities, that is, the methods
   agree upon the metallicity differences between galaxies.
   However, when it comes to absolute [O/H] values, systematic differences
   of up to 0.5~dex between methods can be observed.
   All methods apart from \citet[$x$ axis]{pilyugina,pilyuginb,pilyuginthuan}
   result in unrealistically high gas metallicities for HII galaxies.
   Especially the \citet{kewleydopita,kubilnickykewley} ones
   hardly ever give metallicities below -0.5.
   On top of that, the
   \citet[{[NII]} and ``new paramatrisation'']{kewleydopita,kubilnickykewley}
   methods did not always converge.
   The horizontal lines in the \citet{kewleydopita} panel are due to some
   cases, in which the expression
   $\log(O/H)+12=\frac{-k_1+\sqrt{k_1^2-4k_2(k_0-R_{23})}}{2k_2}$
   gave complex values.
   In these cases, we used only the real part, $-k_1/2k_2$, which
   results in identical metallicity values for galaxies within the same
   ionisation parameter range.
   When ignoring these galaxies, we obtain a similar relation as for
   the \citet{kubilnickykewley} ``new paramatrisation of the
   \citet{kewleydopita}'' method. \\
   The method that yields the most realistic metallicities for the
   galaxies of our sample is the one by \citet{pilyuginthuan}.
   It also accounts for the hardness of the ionising radiation by means
   of the P parameter.
   Therefore, we adopted it for our gas metallicity determination.
   To determine, on which branch a given galaxy lies, we used the
   metallicity estimate based on the
   ${\rm[NII]}~\lambda6548~~\rm{\AA}/{\rm H}\alpha$ ratio by
   \citet*{denicolo}, as suggested by various authors
   \citep[][and others]{perezmonterob,hoyos}.
   Most galaxies of our sample lie on the high-metallicity branch. \\
   \citet{lopezsanchez} come to the same conclusion after comparing
   a series of empirical methods, like the ones presented in this section,
   with the results of the electron temperature method for a sample
   of 31 Wolf-Rayet galaxies.
   They conclude, that the nowadays best suitable method for star-forming
   galaxies, where auroral lines such as [OIII]~$\lambda$4363~\AA\ are not observed,
   is \citet{pilyuginthuan}, and that other methods based on
   photionisation models yield metallicities systematically 0.2-0.3 dex higher
   and higher dispersion than the \citet{pilyuginthuan} calibration.

\subsection{Determination of the stellar metallicities}
\label{FeHstars}
   Then, we determined the metallicity of the stellar content,
   that is, of the stellar populations making up the galaxies,
   using a population synthesis method, described in
   \citet{cuisinier_06} and \citet{lisker}, based on the full spectral
   fitting in the 3820-8570~\AA\ range whereas the upper limit can be
   lower than 8570~\AA\ (in some rare cases as low as 7200~\AA), where the
   deredshifted SDSS spectra do not extend up to this value. \\
   As opposed to the method used for the determination of the line
   intensities, for this population synthesis we used SSP
   spectra from the so-called ``BC99'' SSP library
   \citep{westera_04,cuisinier_06,lisker}.
   It was produced using the GISSEL code,
   implementing the Padova 2000 isochrones
   \citep{girardi_00} combined with the BaSeL 3.1
   ``Padova 2000'' stellar library \citep{paperiii,diss}.
   The BaSeL 3.1 library was calibrated to reproduce the spectral shapes
   of stars of metallicities [Fe/H] from -2 to 0, so it is the ideal
   choice for full spectral fitting.
   By contrast, the ``BC03'' library described in the previous section
   would not have been a good choice,
   for the reasons mentioned there.
   As the GISSEL spectra do not include nebular continuum emission,
   we added it to the spectra, in the same way as described in
   \citet{westera_04,cuisinier_06,lisker}.

   In \citet{westera_04}, we found that the stellar content of
   HII galaxies is made up of at least three populations,
   a young one (up to 10 Myr old) that is responsible for the
   ionisation of the gas and, therefore, for the nebular emission,
   an intermediate one (from 20 to 500 Myr), and an old one
   (at least 1 Gyr old).
   We showed that these three populations are necessary -- and sufficient --
   to characterise an HII galaxy, which is now the current view
   \citep{hoyos}.
   Therefore, in this work
   we modelled the actual population as being composed of an old, an
   intermediate, and a young stellar population.
   In a full spectral fit, we determined the masses, ages and
   metallicities of these partial populations.
   To obtain meaningful results, we chose to reduce the number
   of fitting parameters (and thus, of degeneracies) to a minimum. \\
   As the young population should be more or less coeval with the gas,
   whose metallicity is the one attained now by these galaxies, we
   assumed the young and intermediate populations to have the same
   chemical composition as the gas.
   As the gas metallicities we measured are in terms of [O/H],
   whereas our stellar population library was calibrated in [Fe/H],
   we still had to translate the oxygen - into iron abundances,
   since the element ratios in HII galaxies are far from solar.
   Oxygen is released by stellar winds and by SNe I and II,
   whereas iron and other iron peak elements are only produced in
   SNeIa, which appear only around 1~Gyr after the formation of a
   stellar population.
   Therefore, the ratio between the abundances of these two elements
   in a galaxy depends on its entire star formation history.
   Adding to this the possibility of ``selective mass loss'',
   as is predicted by the models mentioned in the introduction,
   it would be illusory to expect a unique iron-to-oxygen ratio for
   the galaxies of our sample.
   However, as our sample is very homogeneous, we do expect there
   to exist a mean relation and only a moderate scatter around this
   relation.
   Unfortunately, transformations between the two metallicity
   indicators for HII galaxies are rare in the literature.
   In earlier abundance studies of this type of galaxies,
   such as \citet{kniazev} and \citet{izotov}, the iron abundance
   is not determined, as they are also based on emission lines
   in SDSS spectra.
   On the other hand, most theoretical transformations, like the ones
   determined for the Milky Way bulge and the solar neighbourhood by
   \cite{matteucci}, are not valid for HII galaxies.
   Finally, we derived the [O/Fe] to [Fe/H] relation from the one
   for irregular galaxies from \citet{calura}, who calculated the
   element ratios of galaxies of different morphological types using
   chemical evolution models.
   According to their fig.~13, the [O/Fe] to [Fe/H] relation for
   irregular galaxies is around
   ${\rm[O/Fe]}=-0.25\times{\rm[Fe/H]}-0.425\pm0.15$
   independent of redshift, which translates to
   ${\rm[Fe/H]}=1.333\times{\rm[O/H]}+0.5666\pm0.2$.
   Hence, we fixed the metallicities of the young and intermediate
   populations, [Fe/H]$_{\rm y}$ and [Fe/H]$_{\rm i}$, to
   \begin{equation}
    {\rm[Fe/H]}_{\rm y}={\rm[Fe/H]}_{\rm i}=1.333\times{\rm[O/H]}_{\rm P}+0.5666.
   \end{equation}
   For the galaxies of our sample, most of which have
   [O/H]$_{\rm P}$ values in the range from -1 to -0.5, we obtain
   [O/Fe] ratios between -0.2 and -0.4.
   As opposed to the metallicity of the young and intermediate
   populations, the metallicity of the old population, [Fe/H]$_{\rm o}$,
   is a free parameter of our fitting procedure, limited to the
   metallicities of the spectra of the ``BC99'' SSP library, 
   -2.252 -1.65, -1.25, -0.65, -0.35, 0.027, and 0.225.
   Contrary to our approach in previous works, we did not force the
   metallicities of the old populations to be lower than the ones
   of the younger stars, since one of the aims of this work is to
   verify if they really are.

   As the decomposition of galaxy spectra into SSP spectra is known
   to present various degeneracies, we verified if our fitting
   procedure is able to recover the population parameters of synthetic
   composite spectra made up of three SSPs with noise added.
   We found that the masses, ages, and metallicities of the
   input populations were properly recovered for the signal-to-noise
   ratios of both our HII galaxy sample and our test sample
   ($\geq 8$ resp. 5 in the $g$ band), and conclude
   that the number of free parameters of our procedure is adequate
   and that the results presented in the following can be trusted to
   be meaningful.

   In order to determine the total stellar masses of our galaxies,
   we had to know their luminosity distances $D_{\rm L}$, and their
   (stellar) mass-to-light ratios.
   We calculated the former from the redshifts given by the SDSS,
   using a cosmology of $\Omega_{\rm m}=0.3$, $\Omega_{\Lambda}=0.7$,
   and $H_0=70~\frac{{\rm km}}{{\rm s}\cdot{\rm Mpc}}$.
   The stellar mass-to-light ratios result from our best fits.

\section{Results}
\label{results}

   \begin{figure}
    \includegraphics[width=\columnwidth]{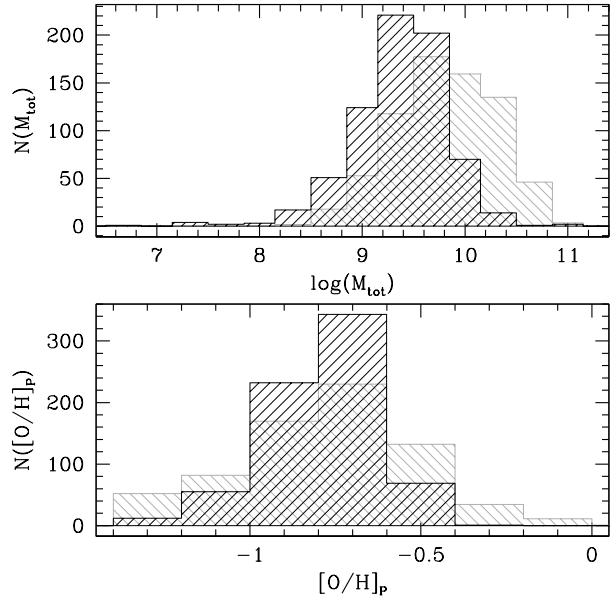}
    \caption{Stellar mass (top panel) - and gas metallicity (bottom panel)
       distributions of the galaxies of our sample (in black).
       The mass - and gas metallicity distributions of the test
       sample are shown in grey (scaled to the same total number
       of galaxies).}
    \label{histos}
   \end{figure}
   \begin{figure}
    \includegraphics[width=\columnwidth]{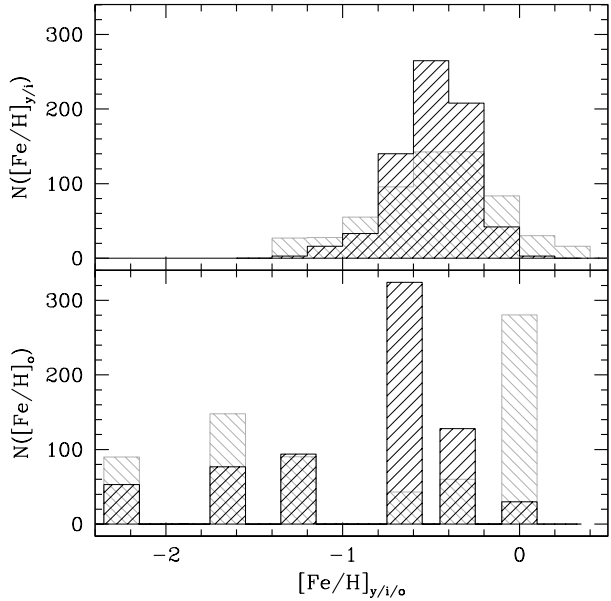}
    \caption{Metallicity distributions of the partial
       populations in our galaxy sample.
       Top panel: the joint young and intermediate age populations,
       bottom panel: old populations.
       The metallicity distributions of the partial populations in the
       test sample are shown in grey (scaled to the same total number
       of galaxies).}
    \label{Fhistos}
   \end{figure}

   The strengths of the emission lines used for this analysis, as well as
   the internal reddening values $E(B-V)$, the gas metallicities
   [O/H]$_{{\rm P}}$ and the total stellar masses $M_{\rm tot}$,
   can be found in Table~\ref{linez} of the electronic version of
   this paper.

   Fig.~\ref{histos} shows the stellar mass and gas metallicity
   distributions of our sample (in black) and the test sample (in grey).
   The average of the logarithms of the stellar masses (in $M_{\odot}$)
   amounts to 9.4 and the average gas metallicity to -0.79,
   whereas those values are 9.8 resp. -0.78 for the test sample. \\
   As expected, the galaxies of the HII galaxy sample have, on average,
   lower masses than the ones of the test sample.
   Since our sample contains very few galaxies with stellar masses
   below $2\cdot10^8~M_{\odot}$ or above $2\cdot10^{10}~M_{\odot}$
   (The mass bins at $5\cdot10^6~M_{\odot}$, $1\cdot10^7~M_{\odot}$,
   $2\cdot10^7~M_{\odot}$, $5\cdot10^7~M_{\odot}$, $1\cdot10^8~M_{\odot}$;
   $5\cdot10^{10}~M_{\odot}$, $1\cdot10^{11}~M_{\odot}$,
   $2\cdot10^{11}~M_{\odot}$, $5\cdot10^{11}~M_{\odot}$,
   and $1\cdot10^{12}~M_{\odot}$ contain 1, 0, 4, 2, 3; 1, 2, 0, 0
   and 0 galaxies, resp.),
   statistical studies of galaxies in these mass ranges suffer from
   low number statistics.
   In the following we will only consider the range
   from $2\cdot10^8~M_{\odot}$ to $2\cdot10^{10}~M_{\odot}$ when studying
   galaxy properties as a function of mass.
   In Figs.~\ref{MtotOH} and~\ref{MtotdF}, this mass range is delimited
   by dotted lines.
   For the test sample, the mass range
   from $2\cdot10^8~M_{\odot}$ to $1\cdot10^{11}~M_{\odot}$ could be used
   for such studies, as each mass bin contains over 150 galaxies. \\
   The average gas metallicities are nearly the same for both samples,
   but the distribution is narrower for the HII galaxies, indicating that
   this is a more homegeneous sample.

   Fig.~\ref{Fhistos} shows the metallicity distributions of the
   partial populations.
   As we fixed ${\rm [Fe/H]}_{\rm y}={\rm [Fe/H]}_{\rm i}=1.333\times{\rm[O/H]}_{\rm P}+0.5666$
   in the fitting procedure, we show in the upper panel the metallicity
   of the joint young and intermediate age population.
   The average metallicity of the joint young and intermediate populations
   is -0.49, and of the old one -0.87
   (test sample: -0.49, and -0.84).
   Here too, the distributions are narrower for the more homegeneous
   HII galaxies sample.
   Nevertheless, the spread in ${\rm [Fe/H]}_{\rm o}$ is wide.
   As ${\rm [Fe/H]}_{\rm o}$ is a measure of the average metallicity
   of all populations older than $\sim 1$~Gyr, this wide spread reflects
   the various stages of (metallicity) evolution of the galaxies
   of our sample around one Gyr ago.

\section{Discussion}
\label{discussion}

   \begin{figure}
    \includegraphics[width=\columnwidth]{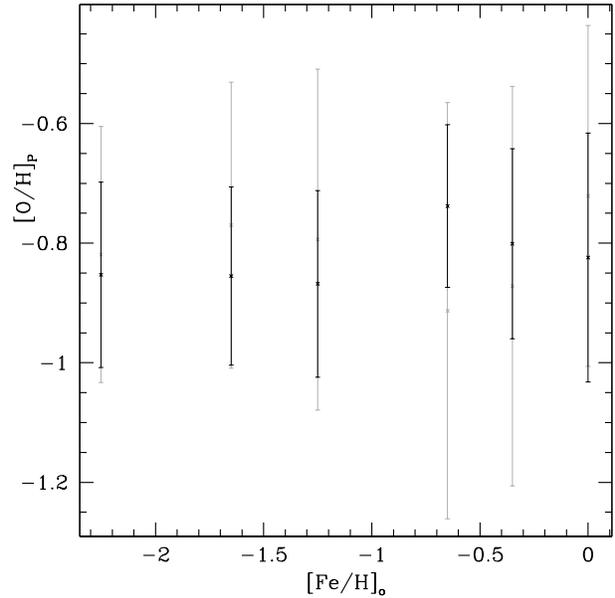}
    \caption{Gas metallicities as a function of the metallicities
       of the old populations.
       The crosses represent the average of the gas metallicities
       for each old population metallicity value, and the error bars
       show the standard deviations.}
    \label{FovsOHp}
   \end{figure}
   \begin{figure}
    \includegraphics[width=\columnwidth]{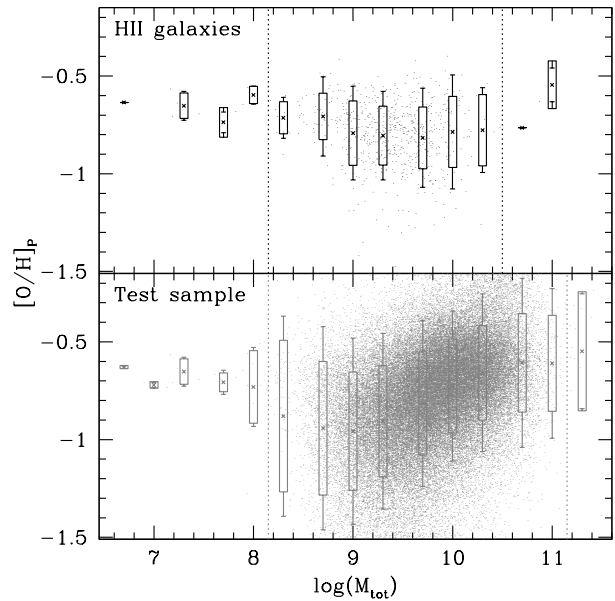}
    \caption{Top panel: the gas metallicity [O/H]$_{\rm P}$ as a function of the
       total stellar mass of the galaxies of our sample.
       The dots show the individual galaxies,
       whereas the crosses, boxes, and bars show the statistical properties
       after subdividing the sample into mass bins.
       For each value of $M_{\rm tot}$, the crosses represent
       the average, the boxes represent the standard deviation,
       and the vertical bars represent the 90th percentile range.
       The dotted vertical lines detach the ``trustworthy'' range
       from the nearly empty mass bins.
       Bottom panel: the individual galaxies, average, standard
       deviation, and 90th percentile range of [O/H]$_{\rm P}$ for
       the test sample.}
    \label{MtotOH}
   \end{figure}
   \begin{figure}
    \includegraphics[width=\columnwidth]{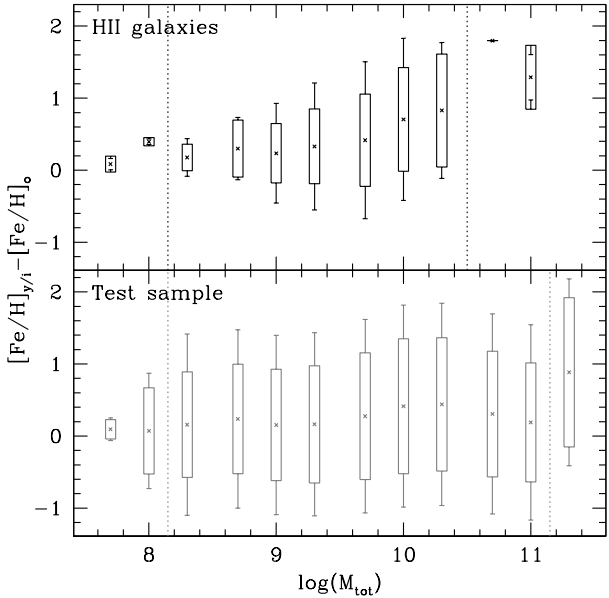}
    \caption{Top panel: the metal enrichment,
       ${\rm [Fe/H]}_{\rm y/i}-{\rm [Fe/H]}_{\rm o}$, as
       a function of the total mass of the galaxies of our sample.
       For each value of $M_{\rm tot}$, the crosses represent
       the average, the boxes represent the standard deviation,
       and the vertical bars represent the 90th percentile range.
       The dotted vertical lines detach the ``trustworthy'' range
       from the nearly empty mass bins.
       Bottom panel: the average, standard deviation, and
       90th percentile range of ${\rm [Fe/H]}_{\rm y/i}-{\rm [Fe/H]}_{\rm o}$
       for the test sample.}
    \label{MtotdF}
   \end{figure}

   In Fig.~\ref{FovsOHp}, we show the gas metallicity in function
   of the metallicity of the old population, both for the HII galaxies
   and for the test sample.
   In either sample, not much of a relation can be seen, any possible
   trend being significantly smaller than the scatter.
   Apparently, the stage of a galaxy's evolution around one Gyr ago
   is not necessarily reflected in the present day one.
   The chemical evolution and the present day (gas) metallicities
   of galaxies suffer stronger influences from other factors, such as the
   galaxy mass.

   Fig.~\ref{MtotOH} shows, how the present gas metallicity [O/H]$_{\rm P}$
   depends on the total stellar galaxy mass.
   In the test sample, [O/H]$_{\rm P}$ clearly increases with mass
   in the range from $2\cdot10^8~M_{\odot}$ to $1\cdot10^{11}~M_{\odot}$,
   as has been found in previous studies
   \citep*[][and others]{tremonti,asari,thomas,panter}.
   However, our mass-metallicity relation lies
   lower than the ones found in these works, as
   recent studies \citep{ellison,mannucci} suggest to be the case for
   samples of star forming galaxies, such as our test sample.

   The HII galaxy sample represents galaxies of a more homogeneous
   class than the test sample.
   Therefore, in this sample, the mass-metallicity relation is
   much less pronounced, if existing at all.
   In fact, the average gas metallicity stays constant
   between -0.7 and -0.8
   in the full galaxy mass range from $2\cdot10^8~M_{\odot}$
   to $2\cdot10^{10}~M_{\odot}$.
   \citet{panter} find almost the same relation for their sub-sample
   of galaxies that are (flux-)dominated by young populations
   ($\leq0.5$~Gyr), as can be seen in their fig.~10. \\
   In a ``closed box'' context, this would mean that the gas fraction
   that has been transformed into stars so far is in the same range for
   all galaxies of the HII galaxy sample, independent of their masses.
   However, HII galaxies are not ``closed boxes'', and our results
   should be interpreted in the light of models including interaction
   processes with the environment, like the ones presented in the
   introduction. \\
   Most of these models predict that HII galaxies lose a large fraction
   of their heavy elements through galactic winds, whose escape
   efficiencies depend upon the galaxy mass.
   These models are at odds with our result for two reasons:
   They result in a dependence of the metallicity on the galaxy mass,
   and their predicted present day metallicities are lower than
   the ones we measure. \\
   The \citet{recchihenslera,recchihenslerb} models with constantly
   infalling clouds appear more promising, especially the models
   NGC-4+BC of \citet{recchihenslera} and NCSM of \citet{recchihenslerb}.
   The clouds partially retain the metal-rich gas, resulting in higher
   present-day metallicities.
   Depending on the star formation history and on the mass of the infalling
   clouds, galaxies like the ones described in these models can reach
   the metallicities we measure.
   On top of that, this scenario provides a possible explanation, why we
   find metallicities in the same ranges for different stellar masses.
   Not only the SFR and, therefore, the amount of produced metals and
   galactic wind increases with increasing galaxy mass, but also the
   amount of infalling low-metallicity clouds and of retained outflowing
   high-metallicity gas.
   Of course, only detailed model calculations can tell, if the mass
   dependencies of the different metallicity-increasing and -decreasing
   processes really add up in such a way, that the final metallicity
   becomes mass independent.
   The fact, that a constant mass-metallicity relation is observed only
   in our HII galaxy sample, and not, for example, in our test sample,
   or in any of the studies mentioned at the beginning of this section,
   shows that only in galaxies with extreme star formation, and thus
   extreme stellar winds, the metal outflow manages to balance the
   retention in this way.
   However, the models including infalling clouds seem to be on the right
   track. \\
   Another noteworthy point of Fig.~\ref{MtotOH} is that, though the
   average gas metallicity [O/H]$_{\rm P}$ does not vary with the galaxy mass,
   its dispersion does increase with galactic mass.
   This indicates that high mass galaxies have a more complex chemical,
   and certainly assembly, history.

   Apart from the present-day metallicity, our analysis also allows us
   to quantify the chemical evolution of our sample galaxies
   since the formation of the old population.
   As the metallicity of the young (and intermediate) population represents
   the galaxy's present-day metallicity, and the one of the old population
   is a measure of the average metallicity of all stars older than
   $\sim 1$~Gyr, the quantity ${\rm [Fe/H]}_{\rm y/i}-{\rm [Fe/H]}_{\rm o}$,
   which we shall call the metal enrichment, is a good measure of the
   change in metallicity during the last few Gyr.
   Fig.~\ref{MtotdF} shows the metal enrichment in function of the
   stellar mass
   for the galaxies in both our HII galaxies sample and the test sample.
   For our HII galaxies, the enrichment is fairly constant
   of the order of 0.3 up to $5\cdot10^9~M_{\odot}$,
   and almost 0.5~dex higher for the last two mass bins within the
   mass range, where reliable statistics are possible,
   $1\cdot10^{10}~M_{\odot}$ and $2\cdot10^{10}~M_{\odot}$.
   This suggests that the enrichment mechanism are independent
   of mass up to $5\cdot10^9~M_{\odot}$, which is compatible with the
   formation scenario favoured in the previous paragraph.
   For galaxy masses above $5\cdot10^9~M_{\odot}$, the enrichment
   is significantly higher, which means that at high masses, the
   metal-rich gas could not escape the galaxy's potential well,
   as predicted in the \citet{maclow} models.
   Here too, the dispersion increases with galaxy mass, confirming
   more complex chemical and assembly histories for high mass galaxies. \\
   An interesting point is that we find a negative metal enrichment,
   thus a metal reduction, for 164 galaxies, that is, 23 per cent of our sample.
   Two mechanisms are known to reduce the (gas) metallicity of a galaxy:
   Selective outflow of metal-enriched gas and infall of low-metallicity
   gas clouds, maybe even of primordial composition.
   Both processes might have contributed to the metal reduction of our
   galaxies, but in the light of our previous results, we expect the
   second one to be more important on long time-scales.

   It is intriguing that in the test sample, too, we find the metal
   enrichment to be mass independent, even though the final metallicity
   increases with galaxy mass.
   ${\rm [Fe/H]}_{\rm y/i}-{\rm [Fe/H]}_{\rm o}$ amounts to around 0.36 in the
   full range from $2\cdot10^8~M_{\odot}$ to $1\cdot10^{11}~M_{\odot}$.
   Even the dispersion is fairly constant over this mass range.
   Apparently, enrichment mechanisms are similar for galaxies of different
   masses even in a heterogeneous sample like our test sample.
   Around 44 per cent of the test sample galaxies, 33178 in total, show
   negative metal enrichment, which proves that processes like selective
   outflow or low metallicity gas infall are also at play in these
   galaxies.

\section{Conclusions}
\label{conclusions}

   We performed a  gas - and stellar population analysis of HII galaxies
   from the Sloan Digital Sky Survey Data Release 7, using their full
   spectra.
   We selected a sample which contains 712 HII galaxy spectra,
   a relatively low number due to our high requirements
   towards the quality of the spectra and to the criterion that the
   spectra contain all the spectral lines necessary for our study.
   We also selected a test sample of 74989 galaxies, using less
   restrictive requirements, the only criterion being that they show
   some line emission.
   Therewith, it is a sample of star forming galaxies. \\
   We derived independent metallicities for the young populations
   present in our sample galaxies from the gas emission lines,
   and for the old populations from full spectra fitting, that is,
   from the continuum and absorption features from stars.

   Our test sample follows the well-known mass-metallicity relation,
   e.g. the fact that the (gas) metallicity increases with (stellar)
   galaxy mass, indicating that low mass galaxies are chemically less
   evolved than high mass galaxies.
   We do not find any systematical tendency of the difference between
   the metallicities of the young and the old stellar populations,
   a quantity we call metallicity enrichment, with galactic mass,
   indicating that, on average, recycling mechanisms of the interstellar
   medium should be the same for low - and high mass galaxies.

   In our HII galaxy sample, on the other hand, we find no mass
   dependence of the present day metallicity.
   For the metallicity enrichment, we find no mass dependence
   for masses up to $5\cdot10^9~M_{\odot}$, whereas above this
   mass, the enrichment is significantly higher.

   We interpret these findings in the light of recent hydrodynamical
   evolutionary models of dwarf irregular galaxies, of which HII
   galaxies are a sub-class.
   We favour models featuring constantly infalling low-metallicity clouds
   able to retain part of the high metallicity galactic winds, such as
   the models NGC-4+BC from \citet{recchihenslera} and NCSM from
   \citet{recchihenslerb}, since models without infalling clouds fail
   to predict the metallicities and the non-dependence on galaxy mass
   of our sample galaxies.
   For galaxies with masses above $5\cdot10^9~M_{\odot}$, 
   most, if not all, of the metal-rich gas is retained by the
   galaxy's gravitational potential, as predicted by \citet{maclow},
   which leads to the observed break-down of the constant mass-enrichment
   relation.

   However, the dispersion of both the present day metallicity and the
   metallicity enrichment does increase with galactic mass,
   indicating more complex chemical and assembly histories for high mass
   HII galaxies. \\
   In 23 per cent of our HII galaxies, the metallicity has decreased in the
   last few Gyr.
   We attribute the metallicity decrease mainly to the dilution of
   the galactic gas by the infalling clouds, which is compatible with
   the models we favour.

\section*{Acknowledgments}

   We would like to thank the Funda\c{c}\~{a}o Carlos Chagas Filho
   de Amparo \`{a} Pesquisa do Estado do Rio de Janeiro (FAPERJ)
   and the PCI programme of ON/MCT (DTI/CNPq) for financial support.
   We would like to thank as well Eduardo Telles
   and Helio Rocha-Pinto for fruitful discussions.
   Finally, the authors acknowledge support by the Swiss National
   Science Foundation.

\appendix

\section{population parameters of individual spectra}

\begin{table*}
      \caption{Line strengths, internal reddening, gas metallicities, and total stellar masses of individual spectra (first 10 lines only, for the complete sample, see the electronic edition).}
      \label{linez}
         \begin{tabular}{lccccccccccc}
            \hline
            \hline
SDSS spectroscopic ID & $I({\rm H}\beta)^*$ & H$\alpha^{**}$ & [OII]$^{**}$ & $S/N_{{\rm [OII]}}$ & [OIII]$^{**}$ & [OIII]$^{**}$ & [NII]$^{**}$ & [NII]$^{**}$ & $E(B-V)$ & [O/H]$_{{\rm P}}$ & $M_{\rm tot}$ \\
MJD-plate-fiber & 4861 \AA & 6563 \AA & 3727 \AA & 3727 \AA & 4959 \AA & 5007 \AA & 6548 \AA & 6584 \AA &  &  & [$10^9~M_{\odot}$] \\
            \hline
51637-0306-583 &  5.102 & 3.010 & 5.848 & 2.948 & 0.728 & 2.088 & 0.148 & 0.472 & 0.088 & -1.022 & 0.63 \\
51658-0282-047 & 10.075 & 3.103 & 5.644 & 2.806 & 0.694 & 2.049 & 0.143 & 0.457 & 0.145 & -0.991 & 1.93 \\
51662-0308-628 & 10.264 & 3.017 & 4.800 & 4.074 & 0.669 & 2.049 & 0.158 & 0.571 & 0.093 & -0.860 & 2.11 \\
51663-0307-268 & 14.076 & 3.052 & 4.270 & 3.005 & 1.009 & 3.078 & 0.176 & 0.574 & 0.114 & -0.794 & 3.27 \\
51671-0299-571 &  8.794 & 3.179 & 6.656 & 2.846 & 0.615 & 2.199 & 0.203 & 0.626 & 0.189 & -1.142 & 6.78 \\
51691-0350-439 & 24.324 & 3.105 & 3.533 & 3.102 & 0.550 & 1.696 & 0.212 & 0.644 & 0.146 & -0.654 & 1.98 \\
51692-0339-437 & 11.115 & 2.921 & 2.707 & 4.272 & 1.238 & 3.741 & 0.074 & 0.223 & 0.033 & -0.599 & 0.34 \\
51783-0395-570 &  9.257 & 2.924 & 3.362 & 4.199 & 0.639 & 1.975 & 0.118 & 0.374 & 0.034 & -0.624 & 0.98 \\
51812-0404-507 &  9.680 & 3.042 & 4.195 & 3.339 & 0.641 & 1.964 & 0.144 & 0.490 & 0.108 & -0.763 & 4.10 \\
51818-0383-266 &  8.743 & 2.944 & 3.631 & 3.398 & 0.899 & 2.789 & 0.097 & 0.303 & 0.047 & -0.687 & 0.87 \\
            \hline
         \end{tabular}
$^*I({\rm H}\beta)$ given in units of 10$^{15}$ erg/s/cm$^2$ \\
$^{**}$flux ratios given in $I(\lambda)/I({\rm H}\beta)$
\end{table*}

\bsp

\label{lastpage}

\end{document}